\documentclass[twocolumn]{aastex63}
\usepackage{amsmath}
\usepackage{enumitem}   
\usepackage{listings}
\usepackage{color}

\newcommand{\msun}{${\rm M_{\sun}}$}

\def\ltsima{$\; \buildrel < \over \sim \;$}
\def\simlt{\lower.5ex\hbox{\ltsima}}
\def\gtsima{$\; \buildrel > \over \sim \;$}
\def\simgt{\lower.5ex\hbox{\gtsima}}
\def\kms{{\rm\,km\,s^{-1}}}

\def\kpc{{\rm\,kpc}}

\def\msun{{\rm\,M_\odot}}

\def\yr{{\rm\,yr}}

\def\g{{\rm\,g}}

\def \r {r$^{1/4}$ }


\def\yr{{\rm yr}}

\def\deg{^\circ}

\def\s{\ifmmode \widetilde \else \~\fi}
\def\={\overline}

\def\spose#1{\hbox to 0pt{#1\hss}}

\def\lta{\mathrel{\spose{\lower 3pt\hbox{$\mathchar"218$}}
     \raise 2.0pt\hbox{$\mathchar"13C$}}}
\def\gta{\mathrel{\spose{\lower 3pt\hbox{$\mathchar"218$}}
     \raise 2.0pt\hbox{$\mathchar"13E$}}}
\def\Dt{\spose{\raise 1.5ex\hbox{\hskip3pt$\mathchar"201$}}}    
\def\dt{\spose{\raise 1.0ex\hbox{\hskip2pt$\mathchar"201$}}}    

\def\r{${\rm r^{1/4}}$}

\def\dotsfill{\leaders\hbox to 1em{\hss.\hss}\hfill}
\def\sun{\odot}
\def\earth{\oplus}

\def\Gyr{{\rm\,Gyr}}
\def\Myr{{\rm\,Myr}}

\usepackage{listings} 
\DeclareFixedFont{\ttb}{T1}{txtt}{bx}{n}{8} 
\DeclareFixedFont{\ttm}{T1}{txtt}{m}{n}{8}  

\usepackage{color}
\definecolor{deepblue}{rgb}{0,0,0.5}
\definecolor{deepred}{rgb}{0.6,0,0}
\definecolor{deepgreen}{rgb}{0,0.5,0}

\definecolor{dkgreen}{rgb}{0,0.6,0}
\definecolor{mauve}{rgb}{0.58,0,0.82}
\definecolor{desertpurple}{rgb}{0.525490,0.176470,0.525490}

\newcommand\pythonstyle{\lstset{
language=Python,
basicstyle=\ttfamily\scriptsize,
numberstyle={\tiny},
otherkeywords={self},             
keywordstyle=\ttb\color{deepblue},
commentstyle=\color{mauve},
emph={MyClass,__init__},          
emphstyle=\ttb\color{deepred},    
stringstyle=\color{deepgreen},
frame=tb,                         
showstringspaces=false            %
}}
\lstnewenvironment{python}[1][]
{
\pythonstyle
\lstset{#1}
}
{}

\def\ltsima{$\; \buildrel < \over \sim \;$}
\def\gtsima{$\; \buildrel > \over \sim \;$}
\def\lsim{\lower.5ex\hbox{\ltsima}}
\def\gsim{\lower.5ex\hbox{\gtsima}}
\def\lapp{\ifmmode\stackrel{<}{_{\sim}}\else$\stackrel{<}{_{\sim}}$\fi}
\def\gapp{\ifmmode\stackrel{>}{_{\sim}}\else$\stackrel{<}{_{\sim}}$\fi}

\DeclareMathOperator{\atantwo}{atan2}

\definecolor{backcolour}{rgb}{0.95, 0.95, 0.96}
\lstdefinestyle{mystyle}{
    backgroundcolor=\color{backcolour},   
    showtabs=false,                  
    tabsize=2
}
\submitjournal{ApJ}

\shorttitle{The \texttt{ACTIONFINDER}}
\shortauthors{Ibata et al.}
\graphicspath{{./}{figures/}}

\begin{document}

\title{The \texttt{ACTIONFINDER}: \\
An unsupervised deep learning algorithm for calculating \\
actions and the acceleration field from a set of orbit segments}

\correspondingauthor{Rodrigo Ibata}
\email{rodrigo.ibata@astro.unistra.fr}

\author[0000-0002-3292-9709]{Rodrigo Ibata}
\affiliation{Universit\'e de Strasbourg, CNRS, Observatoire astronomique de Strasbourg, UMR 7550, F-67000 Strasbourg, France}
\nocollaboration{1}

\author[0000-0002-8788-8174]{Foivos I. Diakogiannis}
\affiliation{ICRAR, The University of Western Australia}
\affiliation{Data61, CSIRO, Kensington, WA 6155, Australia}
\nocollaboration{1}

\author[0000-0003-3180-9825]{Benoit Famaey}
\affiliation{Universit\'e de Strasbourg, CNRS, Observatoire astronomique de Strasbourg, UMR 7550, F-67000 Strasbourg, France}
\nocollaboration{1}

\author[0000-0002-6863-0661]{Giacomo Monari}
\affiliation{Universit\'e de Strasbourg, CNRS, Observatoire astronomique de Strasbourg, UMR 7550, F-67000 Strasbourg, France}
\nocollaboration{1}

\begin{abstract}
We introduce the \texttt{ACTIONFINDER}, a deep learning algorithm designed to transform a sample of phase-space measurements along orbits in a static potential into action and angle coordinates. The algorithm finds the mapping from positions and velocities to actions and angles in an unsupervised way, by using the fact that points along the same orbit have identical actions. Here we present the workings of the method, and test it on simple axisymmetric models, comparing the derived actions to those generated with the Torus Mapping technique. We show that it recovers the Torus actions for halo-type orbits in a realistic model of the Milky Way to $\sim 0.6$\% accuracy with as few as 1024 input phase-space measurements. These actions are much better conserved along orbits than those estimated with the St\"ackel fudge. In our case, the reciprocal mapping from actions and angles to positions and velocities can also be learned. One of the advantages of the \texttt{ACTIONFINDER} is that it does not require the underlying potential to be known in advance, indeed it is designed to return the acceleration field. We expect the algorithm to be useful for analysing the properties of dynamical systems in numerical simulations. However, our ultimate goal with this effort will be to apply it to real stellar streams to recover the Galactic acceleration field in a way that is relatively agnostic about the underlying dark matter properties or the behavior of gravity.
\end{abstract}


\section{Introduction}
\label{sec:Introduction}

Actions and angles are the natural phase-space coordinates to represent the state of the constituents of an integrable, or close to integrable dynamical system, such as typical Galactic potentials \citep{2008gady.book.....B}. In these coordinates, motion along an orbit is very simple, as the actions ${\pmb{J}}$ are preserved, while their canonically conjugate angles ${\pmb{\theta}}$ advance through a cycle at a uniform rate in time (with corresponding fixed frequencies ${\pmb{\Omega}}$). These are also the natural coordinates for perturbation theory \citep[e.g.,][]{Kalnajs1977ApJ...212..637K}. Furthermore, the actions are adiabatic invariants, so slow changes in the evolution of the parent system will preserve these quantities.

With the superb astrometry being gathered by the {\it Gaia} mission \citep{2018A&A...616A...1G}, it has now become possible to study the real six-dimensional phase space structure of our Galaxy in unprecedented detail. Action-angle variables are being used extensively in this exploration, notably (i) because actions can be used to construct equilibrium distribution functions from the Jeans theorem for stars and dark matter \citep[e.g.][]{BinneyPiffl15,Cole17} or for dynamical tracers such as globular clusters \citep{BinneyWong17, Posti19, Vasiliev19}, (ii) because they provide (in principle) the best ``archaeological'' information on the dynamics of the Galaxy \citep[e.g.,][]{Coronado20, Reino21}, and (iii) because they are convenient for perturbation theory. Indeed, after the seminal papers of, e.g., \citet{Dehnen1999,Dehnen2000} -- and more recently \citet{Wegg2015,Portail17} -- on the structure and pattern speed of the Galactic bar, numerous studies have used the action-angle variables to analyze both its linear perturbations and its resonant structure in the Solar neighbourhood \citep[e.g.,][]{Monari17, Binney18, Binney20a, Binney20b, Monari19b, Monari19a, Laporte20, Trick19, Trick20}. 

The present paper represents the first step of a research program aimed at using deep neural networks to perform the canonical transformation to (and from) actions and angles variables. The ultimate goal of our research program will be to apply such methods to galaxy simulations (when close to equilibrium) in a cosmological context as well as to real ({\it Gaia}) data. The present paper, while still limited in scope, is aimed at setting the stage for these future endeavors. Indeed, actions and angles, while undoubtedly useful, are not always easy to calculate. The only potentials in which the actions are expressed analytically are those of from the family of isochrone models of which the Kepler and spherical harmonic potentials are special cases. Although this model is very useful, it does not provide a good approximation to interesting stellar systems (other than those dominated by a central object). Substantial theoretical efforts were therefore undertaken over the past decades to calculate approximations for the transformation from position $\pmb{x}$ and velocity $\pmb{v}$ to $\pmb{J}$ and $\pmb{\theta}$, and conversely, for more general and realistic galactic potentials. \citet{McGill1990} developed a method known as the Torus Mapping \citep[see][for a modern version]{Binney2016}, starting from action-angle coordinates in an isochrone potential, close enough but different from the true potential. Their insight was to search for a generating function in order to transform from the true actions and angles to those of the isochrone. This generating function is expressed as a Fourier series expansion on the isochrone angles, the fitted Fourier coefficients being such that the true Hamiltonian remains, after the associated canonical transformation, constant for a set of true actions. Once this generating function is found, the transformation from actions and angles to positions and velocities is known. For the reciprocal transformation, one usually relies on separable potentials, denoted St\"ackel potentials \citep[e.g.][]{deZeeuw85, Famaey03}, for which three exact integrals of the motion exist. These potentials are best expressed in spheroidal coordinates, associated to a focal distance directly related to the first and second derivatives of the St\"ackel potential. This focal distance can be computed for the true potential at any configuration space point as if the potential was a St\"ackel one, and the corresponding integrals of the motion and respective actions can be evaluated. This method introduced by \citet{Binney12} is known as the ``St\"ackel fudge". Unfortunately, this transformation is not the exact inverse of the one going from actions and angles to positions and velocities as obtained from the Torus Mapping. All these methods are reviewed in \citet{Sanders2016}, and have all been implemented in the {\tt AGAMA} dynamics package \citep[][]{Vasiliev2018,2019MNRAS.482.1525V}.

Our aim in the present work is to provide an innovative way of calculating action-angle coordinates, a method which would be able to jointly determine, from a sample of segments of orbits in positions and velocities, (i) the corresponding actions and angles, as well as (ii) the true acceleration field in which these orbits reside. Moreover, (iii) the inverse transformation from actions and angles to positions and velocities should be determined easily. To this end we will build a deep neural network that will try to learn on its own, and in an unsupervised way, the coordinate transformation from observables into $\pmb{J}$ and $\pmb{\theta}$ for the particular dataset under study. 

Having access to action-angle coordinates can be useful for many types of problems in dynamics: one may wish to study the properties of orbits in exact model potentials, one may wish to track orbits of particles in a frozen or in an evolving N-body simulation, or one may hope to model a real system in nature. Previous approaches to calculating the canonical transformation (reviewed in \citealt{Sanders2016}) have all started by assuming a model potential for the system under study. Thus if the aim of the investigation is to study orbital behaviour in a spherical or axisymmetric (or even sometimes triaxial) potential {\it model}, those previous methods are clearly very well suited to the task, as they directly impose the desired potential on the solutions. In contrast, our method aims at deriving the transformation from $\pmb{x}$ and $\pmb{v}$ to $\pmb{J}$ and $\pmb{\theta}$ in a highly flexible way thanks to a neural network architecture, which should render it useful for the study of systems that do not necessarily obey an analytic density profile or potential imposed upon the system in advance. Our aim with the method was to enable it to find by itself the potential in which the orbits reside (although the solutions can optionally also be forced to be consistent with a pre-selected potential model). We thus expect our algorithm to be applicable to N-body simulations that are either frozen or slowly evolving in time, or for real galaxies or stellar systems in (or close to) dynamical equilibrium. For triaxial systems, especially with strong figure rotation, it is well known that there are many non-regular orbits for which $\pmb{J}$ and $\pmb{\theta}$ are not defined, meaning that our method should be adapted to discard those. Further work will be needed to see if the potential of the system could still be recovered by our method in such cases. We note that, although the method was originally conceived for structures on typical Milky Way halo orbits, the accuracy we obtain for particles on disk-like orbits in a realistic Galactic potential model exceeds the best accuracy expected from the Third Data Release (DR3) of the Gaia mission, a future state-of-the-art dataset.

This contribution is intended as a proof of concept of the method, and to present a novel unsupervised machine learning approach that we expect may be useful to colleagues tackling very different problems in physics.

\begin{figure*}
\begin{center}
\includegraphics[angle=0, viewport= 35 85 890 690, clip, width=16.5cm]{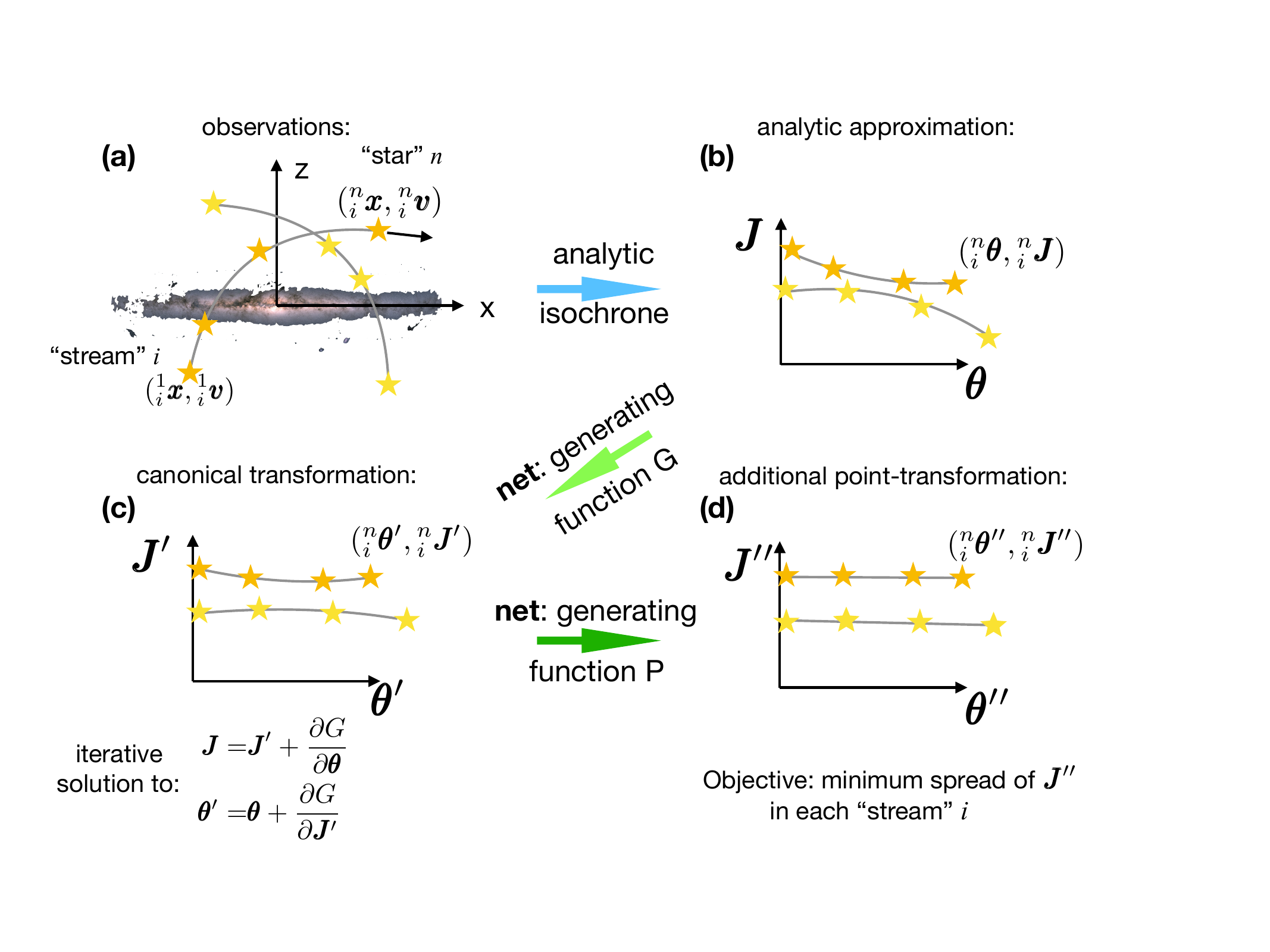}
\end{center}
\caption{Sketch of the method. We consider a dynamical system for which we have measured the kinematics of $N$ particles or ``stars'' along each of a sample of $S$ different orbits which we loosely refer to as ``streams''. In panel (a) we have sketched two such ``streams'', each with four ``stars''. The observable sky positions, parallaxes (or distances) as well as the radial velocities and proper motions are first converted by the algorithm into the Cartesian system $({}_{i}^{n}{\pmb x}, {}_{i}^{n}{\pmb y})$, where the subscript denotes the stream number and the superscript denotes the star number in that stream. The Cartesian coordinates are converted into action-angle coordinates of an analytic isochrone model (b), which serves as a ``toy'' first approximation to the real actions. The central ingredient of the algorithm is the network that proposes trial generating functions $G({\pmb \theta}, {\pmb J'})$ of the canonical transformation between the toy coordinates $({\pmb \theta}, {\pmb J})$ and the refined coordinates $({\pmb \theta'}, {\pmb J'})$. Note that because ${\pmb J'}$ is not known in advance, we need to iterate at this stage (c) to find the ${\pmb J'}$ values that are consistent with the isochrone's values of ${\pmb J}$. Finally (d), we also allow for a simpler canonical point-transformation with the generating function $P({\pmb \theta'})$ to calculate the target coordinates $({\pmb \theta''}, {\pmb J''})$. The two networks (green arrows) are refined over the course of the training procedure by attempting to find the network weights that minimize the spread in the ${\pmb J''}$ values in each stream.} 
\label{fig:sketch}
\end{figure*}

\section{The Algorithm}
\label{sec:Algorithm}

We designed the algorithm around the fact that different phase space points along an orbit have identical $\pmb{J}$ (and frequencies $\pmb{\Omega}$). For this to be useful, we imagine having access to groups of such phase space points along the same orbit. This could be the case if we had the output of a numerical N-body simulation at a series of different time steps, or alternatively we might have such information from observations of a real dynamical system. In the code we refer to a group of such phase space points along the same orbit as a ``stream'', while each individual phase space point is referred to as a ``star''. Our use of the terms ``streams'' and ``stars'' in the software obviously betrays our ultimate motive for building it\footnote{The strategy we intend to explore in future contributions is to fit N-body models to the observed thin and cold stellar streams in the Milky Way using an initial first guess to the global potential. This will allow us to find a mapping between each stream's spatial and kinematic behavior and the orbital track of the corresponding progenitor system. In other words, defining $\phi_1$ to be an angle along the great circle that best fits the stream (with $\phi_2$ orthogonal to $\phi_1$), we will measure the difference between the progenitor orbit and the N-body simulation in distance, sky position, radial velocity and proper motion: $\Delta d(\phi_1)$, $\Delta \phi_2(\phi_1)$, $\Delta v_r(\phi_1)$, $\Delta \mu_{\phi_1}(\phi_1)$, and $\Delta \mu_{\phi_2}(\phi_1)$. These are effectively correction functions for each stellar stream, allowing the positional and kinematic properties of the observed stream stars to be used with the present algorithm in lieu of phase-space coordinates at different times on the progenitor orbit. Once the new acceleration field is found by our neural network, we would then iterate the procedure until the acceleration field derived by \texttt{ACTIONFINDER} is consistent with that used for the N-body models.}, but we note here that star streams do not precisely follow orbits (\citealt{2011MNRAS.413.1852E}; for the dangers of orbit-fitting, see especially \citealt{Sanders2013}), and that a significant amount of further work is required to adapt the present algorithm to properly model such structures.

It is relatively straightforward to devise an algorithm to implement the coordinate mapping from Cartesian $\pmb{\eta} \equiv (x, y, z, v_x, v_y, v_z)$ to $\pmb{\xi} \equiv (\theta_1, \theta_2, \theta_3, J_1, J_2, J_3)$ with a deep neural network for a system for which we have a set of known $\pmb{\eta}$ and $\pmb{\xi}$ values. (This is what we construct for the inverse transformations in Section~\ref{sec:Inverse}). In this case of so-called ``supervised learning'', the network calculates the derivatives of the loss function with respect to the system parameters (the millions of weights in the neural net), and adapts these parameters using a stochastic gradient descent algorithm (in our work we use the ``Adam'' optimization algorithm, \citealt{kingma2017adam}) so as to mimimize the difference between the predicted actions and angles and the corresponding ``ground truth'' values. The learned map can subsequently be used to attribute action and angle values to data not previously seen by the network. However, for the forward transformation in the present work we do not wish to rely on external software to provide us with a training dataset of action-angle coordinates in galaxy models of interest. The fact that nature does not provide us with ground-truth action and angle coordinate labels provides the incentive to be able to calculate these quantities without reference to a training set. The potentials of galaxies probably also do not follow simple mathematical forms, so we have a further motivation to be able to calculate the canonical transformation in a model-independent way.

How then can the stochastic gradient algorithm be steered towards the right solution automatically without labeled data? As stated above, we know that points along the same orbit should have identical actions and frequencies. In addition, the map from $\pmb{\eta}$ to $\pmb{\xi}$ obeys the symplectic condition for the Jacobian matrix:
\begin{equation}
M^T \mathbb{J} M = \mathbb{J} ,
\label{eqn:symplectic}
\end{equation}
where $M$ is the symplectic Jacobian matrix $M_{ij} = {{\partial \eta_i}/{\partial \xi_j}}$, and $\mathbb{J}$ is the antisymmetric block matrix
\begin{equation}
\mathbb{J} = 
\begin{pmatrix}
0 & \mathbb{I}_3\\
-\mathbb{I}_3 & 0 
\end{pmatrix} \, ,
\end{equation}
with $\mathbb{I}_3$ being the $3\times3$ identity matrix. This makes it tempting to simply build a deep network that directly maps $\pmb{\eta}$ to $\pmb{\xi}$, while including the above conditions  as terms in the loss function (together with additional conditions that will be presented in Section~\ref{sec:Network}). We were disappointed to find that this did not work very well. In order to approximate the coordinate transformation to $\sim 1$\% accuracy or better, one requires a deep neural net with tens of millions of parameters. Yet the symplectic condition creates a very complex loss function landscape in this parameter space, which the stochastic gradient descent optimizer explores in a very inefficient way. 

To overcome this problem, we decided to make use of a generating function for the canonical transformation, to guarantee that the coordinate transformation will be symplectic. As pioneered by \citet{McGill1990}, the transformation can be made much simpler by using the action and angle coordinates of a ``toy'' model as a stepping stone. Our procedure is sketched in Fig.~\ref{fig:sketch}. We begin by converting the observed astrometric data into Cartesian coordinates $({\pmb x},{\pmb v})$. It is worth noting that thanks to the automatic differentiation module in the {\it pytorch}  \citep{NEURIPS2019_9015} machine learning library, the gradients of the output quantities with respect to the astrometric observables can be calculated easily; we will make use of this feature in future work to account for observational uncertainties.

For the toy model, we use an isochrone potential:
\begin{equation}
    \Phi(r) = - \frac{G M}{b + \sqrt{b^2+r^2}} \, .
\end{equation}
Here $r$ is the (spherical) radius coordinate, $G$ is the universal gravitational constant, and the two model parameters are the mass $M$ and scale radius $b$. One may then choose the three actions in the isochrone model to be:
\begin{align}
\begin{split}
    J_{1, \rm iso} &= L_z\\
    J_{2, \rm iso} &= L - |L_z|\\
    J_{3, \rm iso} &= {\frac{G M}{\sqrt{-2 E}}} - {\frac{1}{2}} \Bigg( L  + {\frac{1}{2}} \sqrt{L^2 - 4 G M b} \Bigg) \, ,
\end{split}
\end{align}
where $L$ is the angular momentum of the particle, $L_z$ is the $z$-component of angular momentum, and $E$ is the particle's energy. The procedure to convert from $({\pmb x},{\pmb v})$ to the isochrone model's $({\pmb \theta},{\pmb J})$ is detailed in \citet{McGill1990}, and our {\it pytorch} version is heavily inspired by the {\it galpy} isochrone implementation \citep{2015ApJS..216...29B}. 

\citet{McGill1990} found that their algorithm converged irrespective of the initial chosen values of the two isochrone parameters $M$ and $b$, although values closer to the real system produced faster convergence. Our algorithm does not appear to be very sensitive to this initial choice either, as long as $M$ is  set high enough that all orbits are bound. In our algorithm both the $M$ and $b$ parameters can be fitted by {\it pytorch} in the stochastic gradient descent procedure, or if desired, they can be held fixed at their initial values.

We now aim to find new coordinates $({\pmb \theta'},{\pmb J'})$ that are closer to those of the real system. To this end we define an indirect type 2 generating function $G=G({\pmb \theta},{\pmb J'})$, whose derivatives give the implicit transformation:
\begin{align}
    {\pmb J} = \, {\pmb J'} + {\frac{\partial G}{\partial {\pmb \theta}}} \label{eqn:GF_J} \\
    {\pmb \theta'} = \, {\pmb \theta} + {\frac{\partial G}{\partial {\pmb J'}}} \, . \label{eqn:GF_T}
\end{align}
A deep learning network will be used to propose trial variations on $G$, and again thanks to the automatic differentiation in {\it pytorch} it is straightforward to find the ${\pmb J}$ and ${\pmb \theta'}$ values generated by $G$. Since we are actually interested in ${\pmb J'}$, we iteratively find the ${\pmb J'}$ value that yields values of ${\pmb J}$ from Eqn.~\ref{eqn:GF_J} that are the same as those of the toy isochrone.

We now finally update the $({\pmb \theta'},{\pmb J'})$ coordinates using a further transformation ${\pmb \theta''}=P({\pmb \theta'})$ that is only dependent on the updated angles. This point-transformation is much simpler than $G$, as we can obtain the updated angles and actions directly with no need for an iterative procedure,
\begin{align}
    {\pmb \theta''} = \, P({\pmb \theta'}) \label{eqn:GF_Tdd} \\
    {\pmb J''} = \, {\frac{\partial {\pmb \theta'}}{\partial {\pmb \theta''}}} {\pmb J'} \label{eqn:GF_Jdd} \, ,
\end{align}
and is therefore easy for the algorithm to fit. As demonstrated by \citet{1994MNRAS.268.1033K}, such a transformation allows for deformations of the toy model that may be better adapted to the geometry of the real system's orbits. In our experiments we found that this additional network improved the quality of the predicted actions by up to $\sim 50$\%.

Python style pseudocode for the central function of the algorithm is shown in the Appendix in Listing~\ref{list:DECODE}.

\begin{figure}
\begin{center}
\includegraphics[angle=0, viewport= 75 120 550 800, clip, width=\hsize]{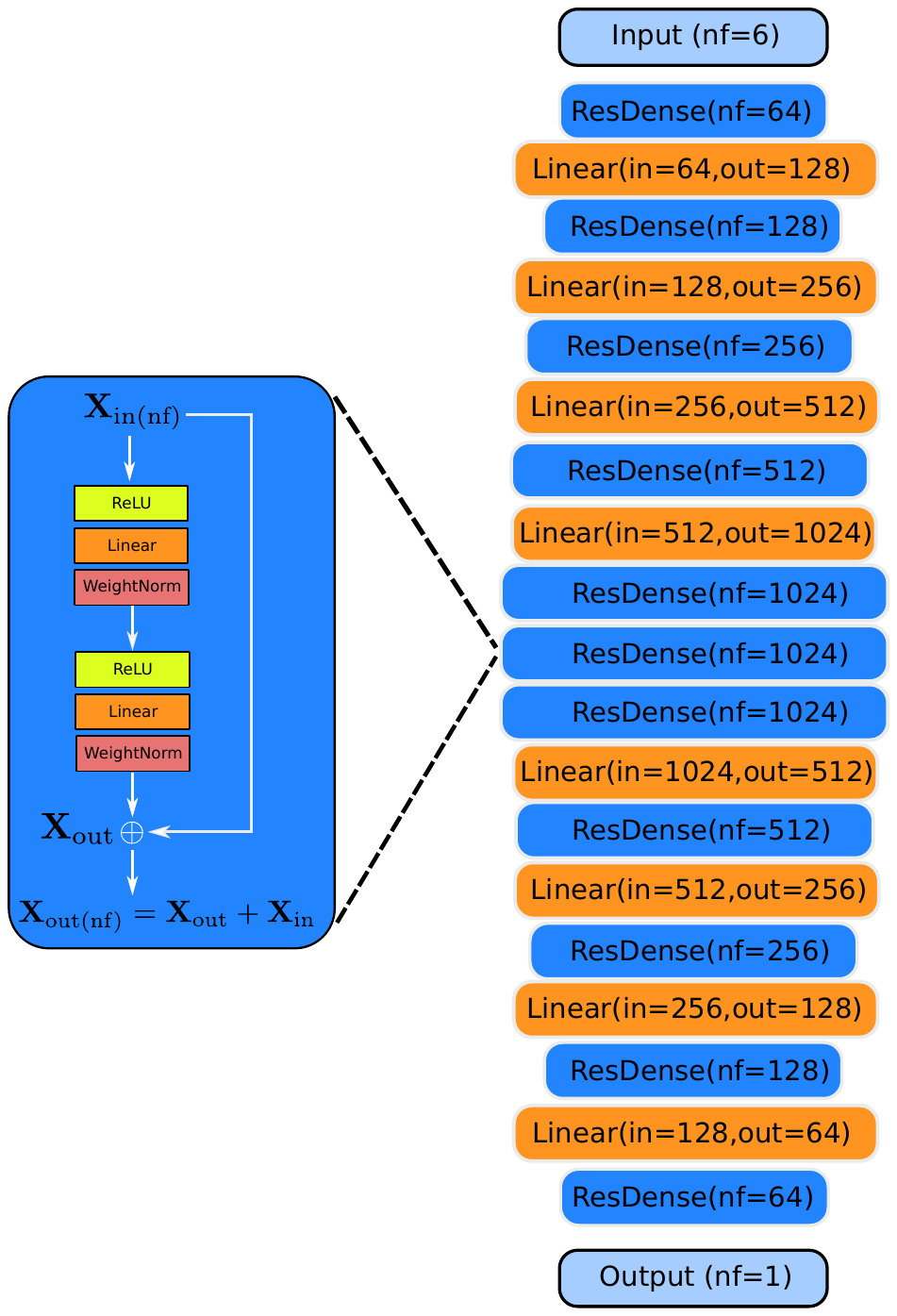}
\end{center}
\caption{Sketch of the (shorter) ${\rm net}_P$ neural network. The ResNet-like unit blocks (left, and Eqn.\ref{eq:basic_block}), are incorporated as a series of layers (right) that progressively increase in complexity up to a chosen maximum width, and then decrease symmetrically up to the layer immediately before the output. The number of features ${\rm nf}$ in each unit block is indicated.} 
\label{fig:architecture}
\end{figure}

\section{The Action-Angle Network}
\label{sec:Network}

All of the neural networks used here have the same basic architecture. Thus, the generating function $G$ is given by:
\begin{align}
    G = {\rm net}_G(\cos({\pmb \theta}),\sin({\pmb \theta}), {\pmb J'}) \, , \label{eqn:net_G}
\end{align}
while the canonical point-transformation $P$ has the form:
\begin{align}
    P = {\rm net}_P(\cos({\pmb \theta}),\sin({\pmb \theta})) \, . \label{eqn:net_P}
\end{align}
After extensive experimentation, we found that the angle coordinates were best learned by first introducing the pair of auxiliary variables $\pmb{t}_{x}\equiv \cos(\pmb{\theta})$ and $\pmb{t}_{y}\equiv \sin(\pmb{\theta})$, as used in Eqns.~\ref{eqn:net_G} and \ref{eqn:net_P}. This naturally takes care of the cyclic property of the angle variables. The ${\rm net}_G$ network consists of a series of 15 blocks similar to a Residual neural network (ResNet, \citealt{2015arXiv151203385H}), adapted for the case of Dense (Linear) layers\footnote{A linear layer $W$ with bias $b$ simply transforms an input vector $x$ to $y=x W^T+b$. The parameters $W$ and $b$ are learned by the algorithm.}. An initial fully-connected linear layer takes the input features (9 in the case of ${\rm net}_G$ and 6 in the case of ${\rm net}_P$) and passes them onto a layer of 64 features. Deeper layers increase in width, becoming a factor of 2  wider in the number of features per layer, up to a maximum of 1024 features per layer. After the chosen maximum depth is reached (at layer 6), the layers decrease in size in a symmetric way. A final fully-connected linear layer is applied to give the chosen output features, which in the case of the generating functions $G$ and $P$ is a scalar value. This architecture is sketched in Fig.~\ref{fig:architecture}. The ${\rm net}_P$ network is identical to ${\rm net}_G$, but consists of only 11 blocks.

The unit blocks were constructed as follows:
\begin{align}
\begin{split}
{\rm out} = W_2(& D_2( {\rm ReLU}( \\
      W_1(& D_1( {\rm ReLU}( {\rm input} ))) \,\, ))) + {\rm input}
\end{split}
\label{eq:basic_block}
\end{align}
where $D_1$ and $D_2$ are fully connected linear layers, ${\rm ReLU}(x)=\max(0,x)$ is the Rectified Linear Unit (ReLU) activation function, and $W_1$ and $W_2$ are Weight Normalisation layers \citep{2016arXiv160207868S}. This setup is very similar to the canonical ResNet, designed so that the blocks learn successive corrections to the input information. We show the pseudo-code of this module in Listing~\ref{list:UNIT_BLOCK} in the Appendix. Each unit block is connected to the following block with a Dense layer, without applying an activation function.

For those unit blocks that exceed a width of 256 nodes, we apply a ``Dropout'' layer \citep{2012arXiv1207.0580H} after every ReLU operation in Eqn.~\ref{eq:basic_block}, to randomly set half of the node weights to zero. This is a commonly-used regularization technique that helps to avoid over-fitting.

Our use of Weight Normalisation layers in Eqn.~\ref{eq:basic_block} is noteworthy. Most modern networks in computer vision make use of the ``Batch Normalization'' (BN, \citealt{2015arXiv150203167I}) procedure to decouple as much as possible the fitting of the parameters in the different layers of a network. This mitigates against the co-variance between layers, allowing the parameters of a deep layer to be refined even though the parameters of the higher layers are also being adjusted at the same time by the optimization algorithm. However, after much experimentation, we found that the BN layers we initially used were limiting the accuracy of our algorithm. This is due to the fact that this procedure operates (during training) on the data presented to it in each ``batch'' (i.e. in small sub-samples which are chosen so that the data may fill the graphics card memory), from which it normalizes the data using the mean and standard deviation of the sub-sample. Since the data at different positions within the batch belong to different streams, the unavoidable shot noise then creates substantial variation between batches, resulting in unacceptably large errors for a study such as ours. We devised a work-around by training in the normal way with the BN procedure until it reached an equilibrium state, and then restarting with the BN layers frozen using parameter values calculated from the whole dataset. We later realised that the ``Weight Normalization'' scheme gives similar accuracy to our BN ``hack'', and have adopted it for the present work as it is an accepted machine learning method.

Between them, with the chosen depth of 15 and 11 layers, the two generating function networks have a total of $\sim 44$~M free parameters.

The chosen loss function is very simple:
\begin{align}
L = L_{J'', \, \rm spread} + L_{J_{2,3}''>0} + \alpha_1 L_{{\theta_0}''} \, .
\label{eqn:loss}
\end{align}
$L_{J'', \, \rm spread}$ is the mean absolute deviation of the difference between the predicted action $\pmb{J''}$ of the stars in a stream and the mean action of that stream $\langle\pmb{J''}\rangle$. The term $L_{J_{2,3}''>0}$ penalises unphysical negative values of ${J_2}''$ and ${J_3}''$, as follows:
\begin{align}
L_{J_{2,3}''>0} = \langle | J_2-|J_2| | \rangle + \langle | J_3-|J_3| | \rangle \, .
\end{align}
Finally, the loss term $\alpha_1 L_{{\theta_0}''}$ is the mean absolute deviation of the target angles ${\pmb{\theta}''}$ when $\pmb{\theta}=0$ is fed into the network (i.e. it encourages the zero-point of the target angles to coincide with those of the isochrone toy model). Following common machine learning practise, we normalised the position and velocity variables (to $20\kpc$ and $200\kms$, respectively); thus the actions are normalized to $4000 \kms \kpc$. However, for convenience, we left the angle variables in radians. We chose to set the hyper-parameter to $\alpha_1=0.1$ in Eqn.~\ref{eqn:loss}, so as compensate for this difference in the range of the angle and action variables.

The algorithm proceeds by iterating over discrete ``epochs'' when the network derivatives are calculated and the parameters are consequently updated to obtain improved estimates for $\pmb{J''}$ and $\pmb{\theta''}$.

\section{The acceleration network}

After the algorithm has converged on values of ${\pmb{J''}}$ and ${\pmb{\theta''}}$ that minimize the spread of ${\pmb{J''}}$ in the streams, we can use these transformed coordinates to study the system they inhabit. To this end, we re-run the network once more with an updated loss function, and with an additional network (the acceleration network) turned on.

The structure of the acceleration network is set up as an additive correction to the acceleration of the toy isochrone model ${\pmb{a}_{\rm iso}}$:
\begin{equation}
\pmb{a} = {\rm net}_{a}(x, y, z, {\pmb{a}_{\rm iso}}) + \pmb{a}_{\rm iso} \, .
\label{eq:net_a}
\end{equation}
Note that here we only feed in the position part of the $\pmb{\eta}$ phase space vector. We use a depth of 15 blocks for this network, which contains $\sim 17.5$~M free parameters. The structure of ${\rm net}_{a}$ is essentially identical to that of ${\rm net}_{G}$, except that it outputs a vector. 

If ${\pmb{J''}}$ and ${\pmb{\theta''}}$ are only functions of position ${\pmb{x}}$ and velocity ${\pmb{v}}$, the chain rule implies that the mapping is constrained by the following relations:
\begin{align}
{\frac{d {J_i}''}{d t}} &= {\frac{\partial {J_i}''}{\partial x_j}} \dot{x}_j + {\frac{\partial {J_i}''}{\partial \dot{x}_j}} \ddot{x}_j = 0
\label{eqn:J_dot}\\
{\frac{d {\theta_i}''}{d t}} &= {\frac{\partial {\theta_i}''}{\partial x_j}} \dot{x}_j + {\frac{\partial {\theta_i}''}{\partial \dot{x}_j}} \ddot{x}_j = {\Omega_i}'' \, ,
\label{eqn:theta_dot}
\end{align}
where the $i$ index here is associated to each of the (usually three) integrals of motion $J_i$, and $j$ is a dummy index denoting Einstein summation. 

A brief inspection of Eqn.~\ref{eqn:J_dot} reveals that once we have access to a network that delivers predictions for the position and velocity derivatives of $J''$, we can solve the simultaneous equations to calculate the acceleration terms $\ddot{x}$. While this suggests that the acceleration network (Eqn.~\ref{eq:net_a}) is superfluous, we nevertheless opted to employ a network to calculate $\pmb{a}$ for two main reasons. First, it allows us to train a network that can be subsequently applied to make predictions for the acceleration on new unseen data (for instance, for ``stars'' without full 6-dimensional information). Second, we found that solving Eqn.~\ref{eqn:J_dot} for $\ddot{x}$ using linear algebra can occasionally be problematic, due to ill-conditioned matrices. Creating a separate acceleration network eliminates these problems. 

In this second run through the algorithm, we choose the following loss function:
\begin{align}
\begin{split}
    L =& L_{dJ''/dt} + \alpha_2 L_{\rm symmetry} + \alpha_3 L_{\Omega'', \, \rm spread} \,  ,
\end{split}
\label{eqn:loss_acc}
\end{align}
where $L_{dJ''/dt}$ is the mean absolute value of $dJ''_i/dt$, as calculated from Eqn.~\ref{eqn:J_dot}. $L_{\rm symmetry}$ is an optional term to enforce a desired symmetry on the solution of the acceleration network. For instance for axisymmetric potential models, we experimented with a cosine anti-similarity criterion $L_{\rm symmetry} = 1 + \vec{a}_R \cdot \vec{R}/(||\vec{a}_R|| \, ||\vec{R}||)$ between the cylindrical-$R$ component of the acceleration field $\vec{a}_R$ and the same component of the (Galactocentric) position vector $\vec{R}$. Although this loss term helps to constrain the acceleration, for the purpose of the tests presented in Section~\ref{sec:Results} we decided to suppress this symmetry constraint on $\pmb{a}$ by setting $\alpha_2=0$. 

Finally, $L_{\Omega'', \, \rm spread}$ is another optional loss term to ensure that the frequencies calculated from Eqn.~\ref{eqn:theta_dot} for each ``stream'' have minimum spread. This is done by calculating the mean absolute difference of the frequency of each ``star'' from the mean of the frequencies in the ``stream''. However, we found in our tests that including this term introduced slight additional scatter in the solutions (most likely because angles are much less accurately constrained than actions in our algorithm), so it too was suppressed by setting $\alpha_3=0$ in the present work.

\section{Inverse Transformation}
\label{sec:Inverse}

Deep learning can also be used to calculate the inverse transformation from actions and angles back to positions and velocities. However, there are many different situations that one can envisage being confronted with. For instance, one may or may not have frequencies in addition to the actions and angles, or one may already know the potential. Alternatively, it is possible that the potential and frequencies are not known, but one may have groups of particles along orbits (i.e. our ``stars’’ in ``streams’’). Each of these cases would require a different network to be constructed in order to learn the inverse transformation in an unsupervised way.

The approach we take here is slightly different, but perhaps more realistic given the forward transformation we have presented previously. We imagine having derived the forward transformation from $\pmb{\eta}$ to $\pmb{\xi}$ for a number of orbits. This information can now be used to train a supervised network to predict $\pmb{\eta}$ from $\pmb{\xi}$. 

Since this is a supervised learning task, compared to the previous problem, it is a much easier for the network to find the correct direction for the stochastic gradient descent to improve the loss function. We therefore choose a simple network architecture:
\begin{equation}
(\pmb{t}_x,\pmb{t}_y,\pmb{J}) = {\rm net}_{\rm inverse}(\pmb{t}_x\pmb{''},\pmb{t}_y\pmb{''},\pmb{J''}) \, ,
\end{equation}
from which the toy model angles are calculated with $\pmb{\theta}=\atantwo(\pmb{t}_y,\pmb{t}_x) + \pi/2$.
Just like the forward network, the inverse transformation thus employs the isochrone model as a stepping stone, with the final step being the analytic inversion of the toy model's $(\pmb{\theta},\pmb{J})$ coordinates to the 6-dimensional vector $\pmb{\eta}$. Since this inverse network is not constrained by differential equations (and hence does not require any Jacobians to be calculated) it is feasible to implement it as a deeper network of 23 (or more if necessary) layers of blocks (Eqn.~\ref{eq:basic_block}), using a total of $\sim 34$~M free parameters.

\section{Results}
\label{sec:Results}

The algorithm was constructed to accept as input data a set of $S$ ``streams'', each with $N$ ``stars''. The method works with $N\ge2$ ``stars'' per ``stream'', but for these initial tests we have chosen a more conservative $N=8$, simply to make it easier for the algorithm to verify that the actions and frequencies calculated for each ``stream'' are constant. {\it Pytorch} achieves its speed by processing the data in parallel, and so it is much more efficient to pass {\it pytorch} ``tensors'' of the same size onto the graphics processing unit (GPU). Because of this, we expect that when the method is applied to real data, it will be convenient to break up streams with a large number of known members into smaller sub-groups (of the same size $N$).

In all the experiments below, we split the input sample into two sets, a training set with 50\% and a test set with 50\% of the sample. Clearly, with real data we would be much more sparing with the fraction allocated to the test set! Typically, we run the training process with a learning rate of $10^{-4}$. In all the tests reported below we iterate for 1024 ``epochs'' even though the training losses often stabilize much earlier. We found that re-running the network with smaller learning rates (which is the standard procedure in machine learning to improve accuracy), resulted in only very marginal improvements to the loss function values. 

Test samples are used in machine learning primarily to verify that the training procedure is not overfitting the data. When this occurs, the loss of the training set continues to improve, while the loss values in the test set (which the algorithm does not see during training) become worse. We simply ignore all further epochs once the algorithm begins to overfit.

Both the data and network are expressed internally as double precision floating-point numbers, as we found that the Jacobian matrices were not always calculated to sufficient precision with single precision numbers, resulting in a network that would not update correctly due to vanishing gradients.

Because of the fact that we find the actions iteratively starting from an initial guess provided by the toy model, the speed of the algorithm depends on how close the toy model is to the target system. But in typical cases using $S=1024$ and $N=8$, the algorithm takes $\sim 3$~hours to complete a training run of 1024 epochs on an {\it NVIDIA GV100} GPU with 32~GB of card memory. Running times on larger datasets should scale approximately linearly with the number of data.

\subsection{Fitting isochrone models}
\label{sec:isochrone_model}

Although we have built the \texttt{ACTIONFINDER} as a series of transformations from the analytic isochrone, it is still worth checking whether the software can fit a sample of orbital points drawn from different models of this family.

To this end we generated orbits within an isochrone potential with $M=3.334\times 10^{11}\msun$ and $b=5\kpc$ (i.e. $M=1.5\times 10^6$ in N-body units where $G=1$ and distances are in $\kpc$ and time is in $\Gyr$), which gives a circular velocity at the Solar neighborhood ($R_\odot=8.122\kpc$, \citealt{2018A&A...615L..15G}) of $v_c(R_\odot)=217\kms$. One of the advantages of the \texttt{ACTIONFINDER} is that we do not need to provide it with a {\it fair} sample of orbits within the system, almost any sample that covers the region of interest will do. 

In the following, we imagine having access to the orbits of some objects in the ``Galaxy'' halo. To model this situation we select a random 3-dimensional initial radius drawn uniformly in Galactocentric distance between $r=[6$--$16]\kpc$, together with a velocity vector that is oriented randomly (i.e. isotropic) with magnitude drawn from a Gaussian of dispersion $150\kms$. Using a symplectic Leapfrog scheme, we integrate from these initial phase space locations for $0.1\Gyr$, ensuring energy conservation to 1 part in $10^7$, and randomly select $N=8$ phase space points along each path. Since our aim is to eventually work with real astrometric data, we convert the positions and velocities of the set of orbit locations to the observable quantities: $d, \ell, b, v_h, \mu_\ell, \mu_b$, as they would be measured from our vantage point in the Galaxy, where $d$ is Heliocentric distance, ($\ell, b$) are Galactic coordinates, $v_h$ is the Heliocentric radial velocity, and $\mu_\ell$ and $\mu_b$ are the proper motions along the Galactic coordinate directions.

With the correct input values of $M$ and $b$, the network quickly finds the correct $\pmb{J''}$ and $\pmb{\theta''}$ to better than 0.01\% (with a training set size of $S=1024$), which is reassuring given that the algorithm effectively just has to learn (in an unsupervised way) the identity operation.

A more interesting case occurs when we try to fit the isochrone ``stream'' sample above using fixed and incorrect reference $M$ and $b$ values. With $M$ and $b$ both 10\% lower (higher) in the toy model compared to the simulated data, using $S=1024$ we obtain action errors of $\delta \pmb{J}=1.2 (2.0) \kms \kpc$ and angle errors of $0.7\deg (1.2\deg)$. When $M$ and $b$ are 20\% higher, the action error is $\delta \pmb{J}=3.0 \kms \kpc$ and the angle error is $1.4\deg$. The algorithm fails with a fixed 20\% lower value of $M$, because some orbits are unbound in the toy model, and thus do not have valid $(\pmb{\theta},\pmb{J})$ coordinate values.

Using $S=1024$ with fixed 20\% higher values of $M$ and $b$ in the toy model, the inverse transformation from $(\pmb{\theta'},\pmb{J'})$ to $(\pmb{x},\pmb{v})$ is recovered to an accuracy of $\delta \pmb{x}=0.3\kpc$ and $\delta \pmb{v}=2.6\kms$ (again, mean absolute deviation errors).

\begin{figure}
\begin{center}
\includegraphics[angle=0, viewport= 25 40 660 650, clip, width=\hsize]{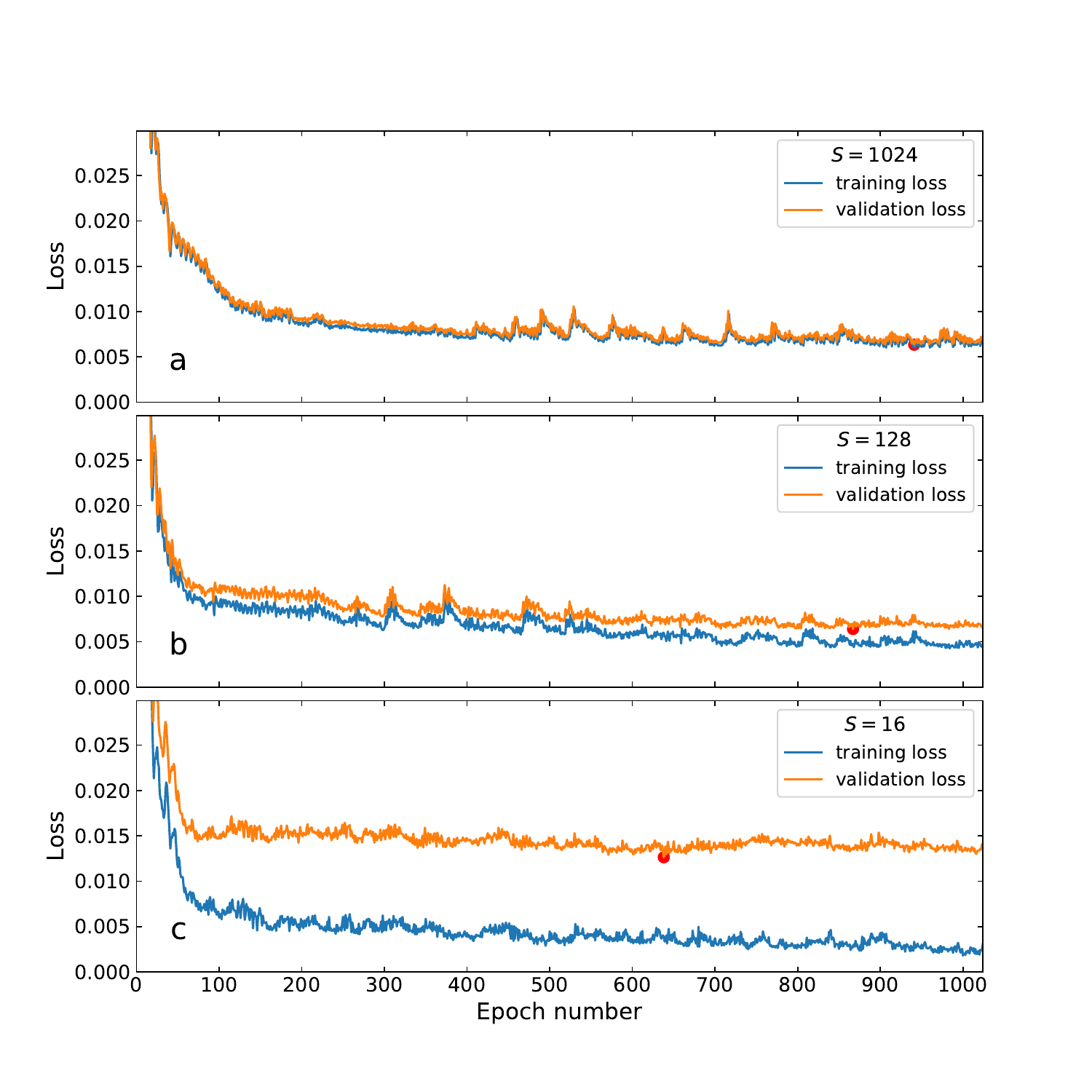}
\end{center}
\caption{Convergence of the loss function for different sample sizes. All three tests use simulated orbits in the \citet{1998MNRAS.294..429D} potential model `1', each with $N=8$ phase-space points. The sample sizes are $S=1024$, $S=128$ and $S=16$ for the top, middle and bottom panels, respectively. In each case, the blue line shows the training loss while the orange line shows the validation loss. The red dot shows the position of the best validation loss, which defines the epoch at which the $(\pmb{\theta''},\pmb{J''})$ coordinates are extracted.} 
\label{fig:loss}
\end{figure}

\subsection{Fitting the Dehnen \& Binney Galactic model}
\label{sec:DB_model}

We now examine the performance of our method with two reasonably realistic Galactic potential models, by \citet{1998MNRAS.294..429D} and \citet{2014MNRAS.445.3133P}, who fitted plausible axisymmetric density models of the main Galactic components to kinematic observations of the Milky Way. These useful models have been incorporated into the {\tt AGAMA} package \citep{2019MNRAS.482.1525V}, which now allows one to transform the actions and angles of test particles in these potentials to positions and velocities with the Torus Mapping. We first consider model `1' of \citet{1998MNRAS.294..429D}, in which we integrated a set of random orbits (each with fixed triplet of actions) in a similar way to that described in Section~\ref{sec:isochrone_model}, and as before we selected $N=8$ random points along the orbits, recording their input actions and angles values, their $(\pmb{x}, \pmb{v})$ values generated with the Torus Mapping, as well as their output $(\pmb{\theta''}, \pmb{J''})$ values found by \texttt{ACTIONFINDER}. For these tests the toy isochrone $M$ and $b$ parameters are left free.

\begin{figure*}
\begin{center}
\vbox{
\hbox{
\includegraphics[angle=0, viewport= 15 40 660 650, clip, width=9cm]{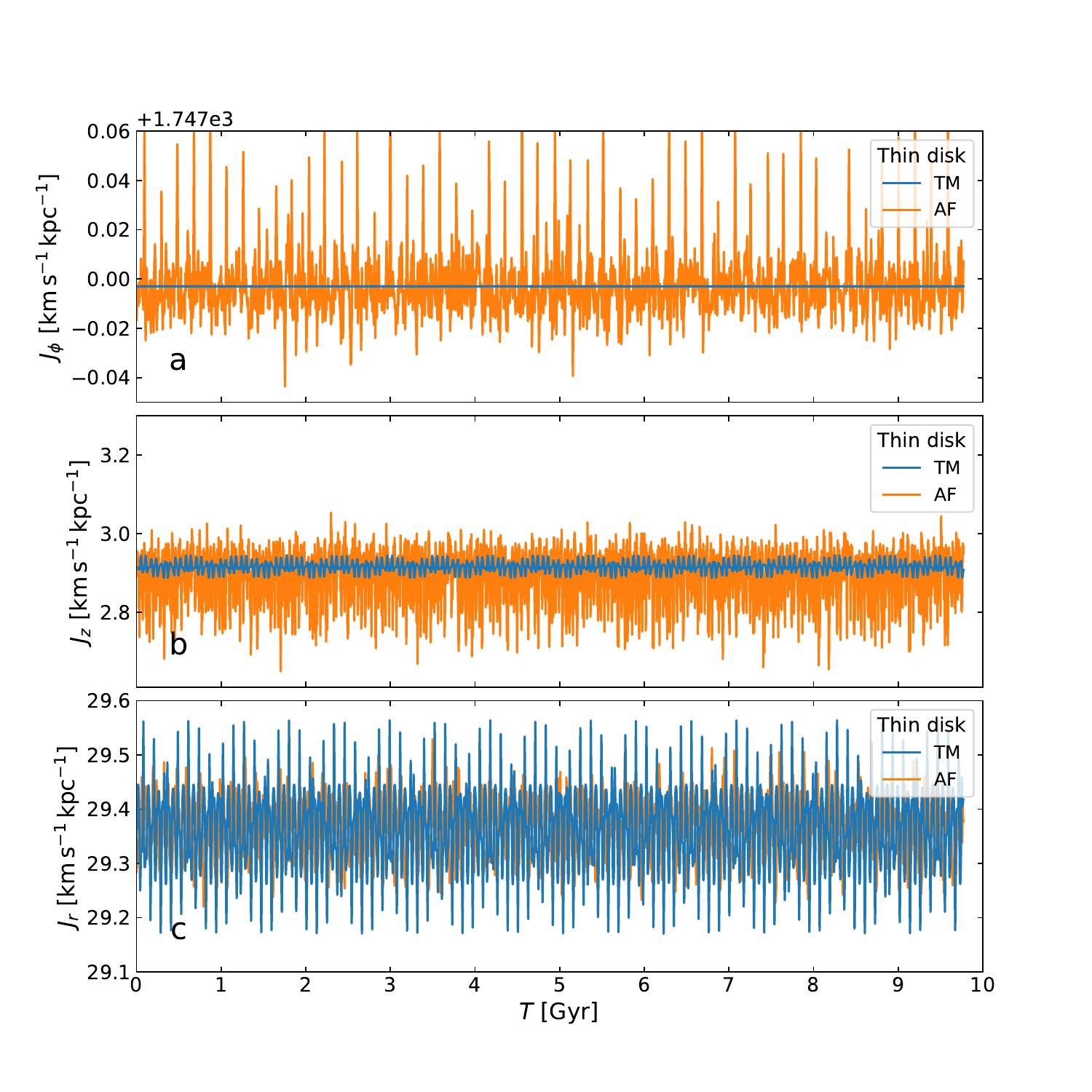}
\includegraphics[angle=0, viewport= 15 40 660 650, clip, width=9cm]{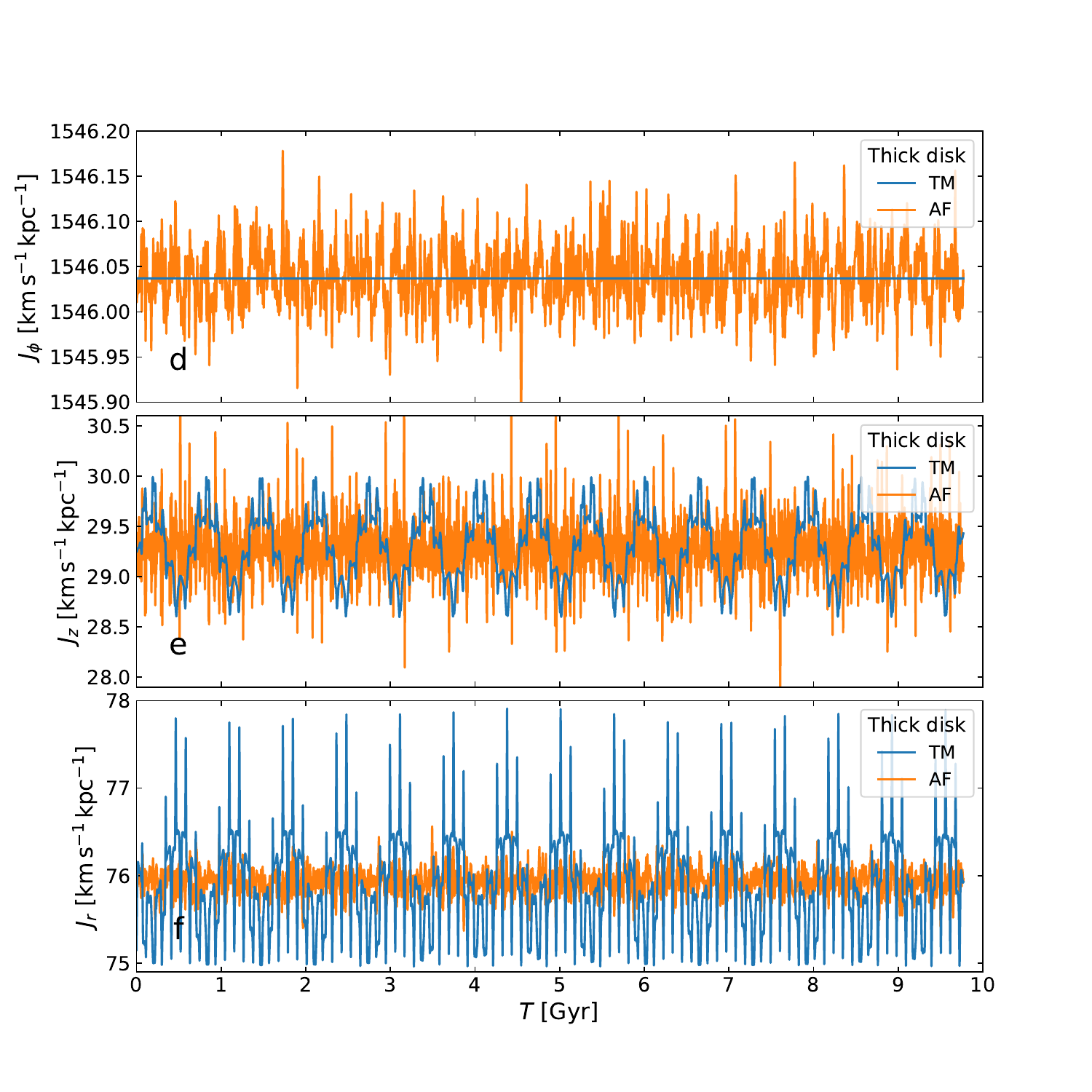}}
\hbox{
\includegraphics[angle=0, viewport= 15 40 660 650, clip, width=9cm]{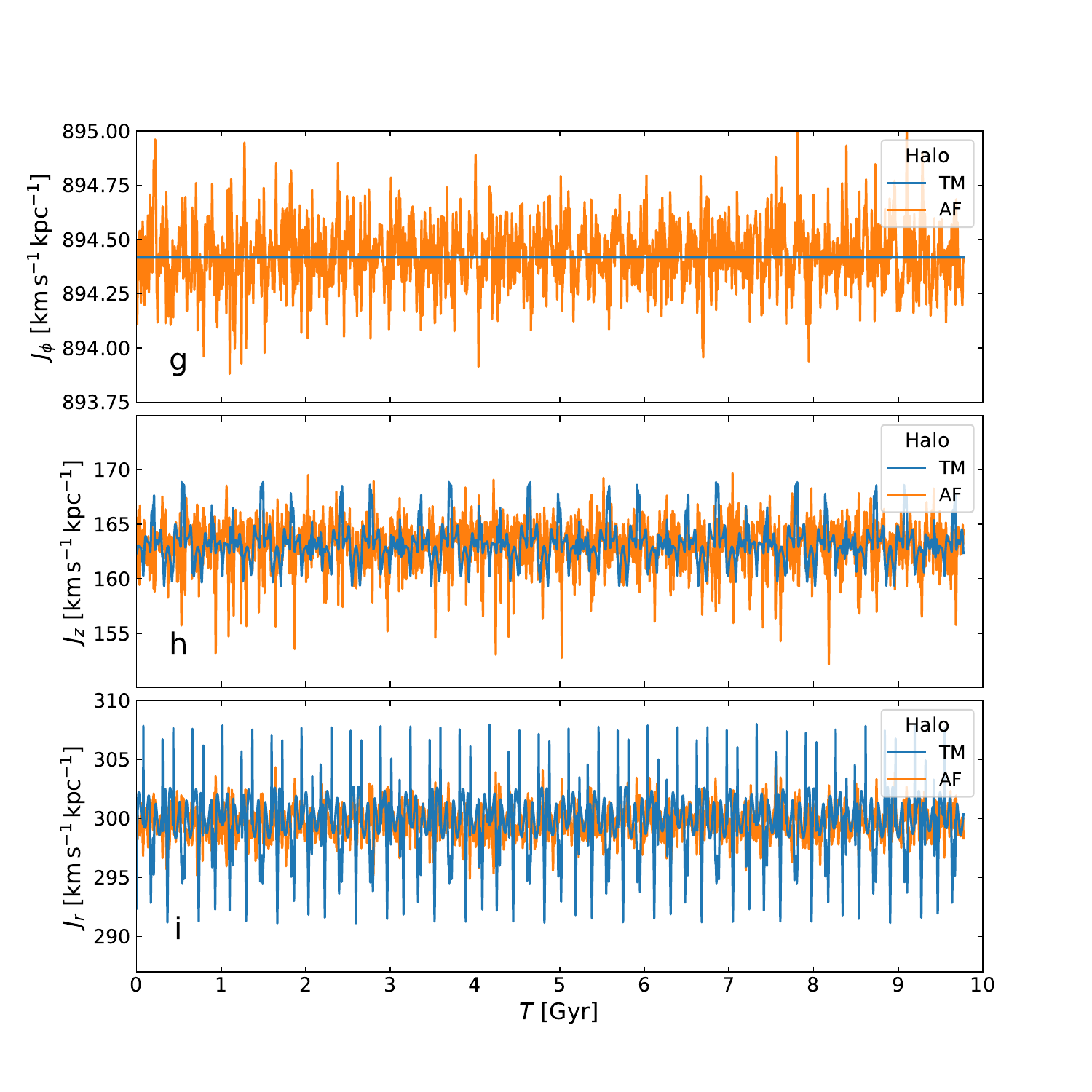}
\includegraphics[angle=0, viewport= 15 40 660 650, clip, width=9cm]{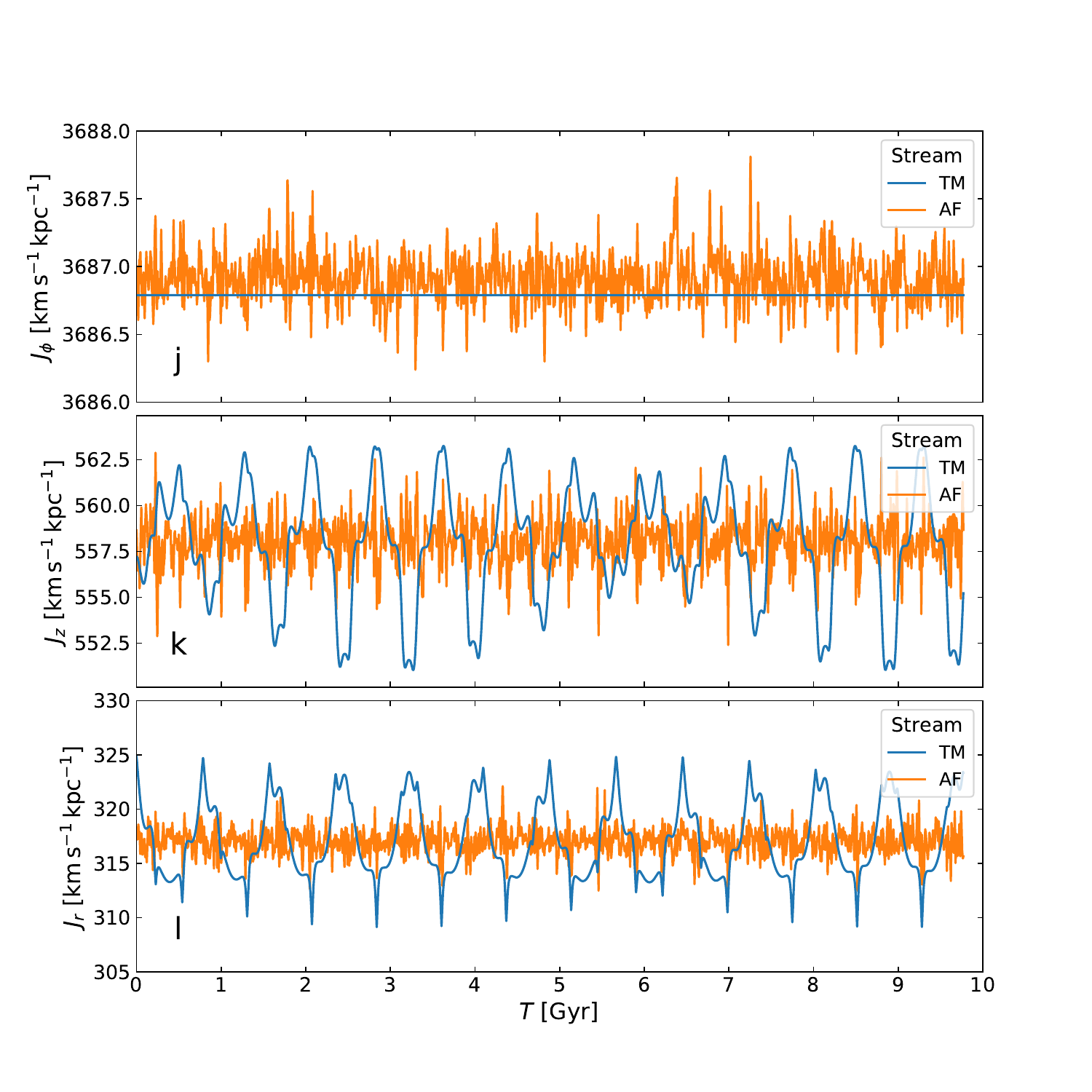}}
}
\end{center}
\caption{Comparison of the action values derived with the Torus Machinery (TM, blue) and the present algorithm (AF, orange) for the four test orbits defined by \citet{Sanders2016} in the \citet{2014MNRAS.445.3133P} potential model. Here, the Deep Learning algorithm is only provided $S=1024$ sets of $N=8$ data points along each orbit, and the network is initiated with random numbers. We include $J_\phi$, since in our case the network needs to discover from the data that the host system is axisymmetric.}
\label{fig:test_orbits}
\end{figure*}

Using $S=1024$, the \texttt{ACTIONFINDER} recovers the Torus Mapping input action-angle coordinates with an error of $\delta \pmb{J''}=8.0\kms \kpc$ in action (i.e. $\sim 0.4$\% of the Sun's action) and $\delta \pmb{\theta''}=7.4\deg$ in angle. The model accelerations are also recovered to $\delta \pmb{a}=2.6$\%. These errors are calculated as the mean absolute deviation between the Torus Mapping input values and the \texttt{ACTIONFINDER} predictions. With these noise-less data (possessing only shot-noise), the derived $(\pmb{\theta''},\pmb{J''})$ are not very sensitive to the sample size. If instead we set $S=128 (16)$, we obtain an action error of $\delta \pmb{J''}=11.2 (18.1)\kms \kpc$ and the angle error is $\delta \pmb{\theta''}=8.0\deg$ ($9.1\deg$). In Fig.~\ref{fig:loss} we show the evolution of the loss function in these three tests.

As a comparison, we also transformed back the positions and velocities of each particle into action-space with the St\"ackel-fudge in {\tt AGAMA}. In this case, the mean absolute deviation between the Torus Mapping input actions and the St\"ackel-fudge estimate is $44 \kms \kpc$, $5$ times less accurate than with the \texttt{ACTIONFINDER}.

With $S=1024$ the inverse transformation algorithm (which is basically the same transformation as the one done with the Torus Mapping to generate positions and velocities) recovers the $(\pmb{x},\pmb{v})$ originally generated with the Torus Mapping to an accuracy of $\delta \pmb{x}=0.24\kpc$ and $\delta \pmb{v}=4.4\kms$.

\subsection{Fitting the Piffl et al. Galactic model}
\label{sec:Piffl_model}

For the second realistic Galactic potential we test our algorithm with, we choose the model of \citet{2014MNRAS.445.3133P}, which was used in the review by \citet{Sanders2016} as the reference potential for comparing different methods for calculating actions. Repeating the selection of halo orbits in an identical manner to what was done above with the \citet{1998MNRAS.294..429D} model results in a similar action error of $\delta \pmb{J''}=6.3\kms \kpc$ and $\delta \pmb{\theta''}=4.2\deg$ with $S=1024$.

When we initially devised the present algorithm, we intended to apply it to objects on halo-type orbits. Nevertheless, it is interesting to consider how well it can perform for stars or particles on disk orbits. To this end, we selected a sample of disk-like orbits in the \citet{2014MNRAS.445.3133P} potential model. To do so, we generate pseudo-random actions $\pmb{J}$ uniformly distributed in the ranges $[0,50\kms\kpc]$ for $J_r$, $[0,50\kms\kpc]$ for $J_z$, and $[0,3000\kms\kpc]$ for $J_\phi$. For each orbit, we then generate 8 angle values $\pmb{\theta}$, where each component is uniformly distributed in the range $[0,2\pi]$. To each action angle position $(\pmb{J},\pmb{\theta})$ we associate positions and velocities $(\pmb{x},\pmb{v})$ obtained with the Torus Mapping technique. We avoided the bulge region by discarding orbits with pericenters smaller than $5\kpc$. Using $S=8192 (1024)$ orbits, we recovered the input actions with an error of $\delta \pmb{J''}=0.67 (2.4)\kms \kpc$, and $\delta \pmb{\theta''}=4.9 (9.9)\deg$ in angle. While performing these tests, we found that we could improve the quality of the recovered actions if we simplified the loss function to 
\begin{align}
L = L_{J'', \, \rm spread} \, ,
\label{eqn:loss_simple}
\end{align}
i.e. by giving up on measuring angle variables with correct zero-points. We were then able to recover the Torus Mapping actions of the same sample of disk-like orbits with an error of $\delta \pmb{J''}=0.088 \kms \kpc$ using $S=8194$ orbits (i.e. a factor of $\sim 8$ improvement in accuracy can easily be obtained when the angle variables are not required).

\subsection{Four test orbits}
\label{sec:test_orbits}

When comparing the different methods for calculating actions, \citet{Sanders2016} also examined how accurately the actions were derived along four test orbits. These tests were undertaken for a typical orbit in the thin disk, the thick disk, the halo, and in the GD-1 stellar stream \citep{2006ApJ...643L..17G}. We now subject our algorithm to these four tests. We integrated the four pre-defined orbits in the \citet{2014MNRAS.445.3133P} potential, storing the output phase space position and Torus Mapping actions at 10,000 intervals each equally separated by $0.9777\Myr$. The algorithm was then provided as input $S=1024$ sets of $N=8$ phase space points chosen randomly along the orbits. After completing the training on this dataset (using the simplified loss function of Eqn~\ref{eqn:loss_simple}), the learned coordinate mapping was applied individually to each of the 10,000 orbital points. The resulting actions thus derived by the algorithm are shown in Figure~\ref{fig:test_orbits} (orange) and are compared to the values derived with the Torus Machinery in {\tt AGAMA} (blue). We include panels for $J_\phi$, contrary to \citet{Sanders2016}, since the algorithm was not forced to find a trivial solution consistent with an axisymmetric potential, but discovered this component of the action from the data themselves. The mean absolute deviation of the actions of the thin disk, thick disk, halo and stream test orbits are $0.034\kms \kpc$, $0.11\kms \kpc$, $0.84\kms \kpc$ and $0.61\kms \kpc$, respectively; while the same statistics using the Torus Machinery in {\tt AGAMA} are $0.023\kms \kpc$, $0.23\kms \kpc$, $0.94\kms \kpc$ and $1.82\kms \kpc$, respectively. We note that significantly lower dispersions can be obtained with our method if instead of initializing the neural network with random numbers, we provide as starting points the solutions previously fitted on the sample of disk or halo orbits discussed in Section~\ref{sec:Piffl_model} (this yields $0.012\kms \kpc$, $0.064\kms \kpc$, $0.47\kms \kpc$ and $0.14\kms \kpc$ for the mean absolute deviation of the actions of the thin disk, thick disk, halo and stream test orbits, respectively).

\subsection{Fitting an N-body simulation}
\label{sec:Nbody}

As we have mentioned above, we also expect our algorithm to be useful for analysing the dynamics N-body systems. For the method to be applicable, the actions of the particles in the system need to be conserved. This implies that the system needs to be non-collisional and in equilibrium, or close to equilibrium; furthermore, the particles to be analysed cannot reside on chaotic orbits. Here we chose as a first example a spherical Plummer model for which we are easily able to calculate ground-truth actions, thus allowing the accuracy of the method to be gauged.

\begin{figure}
\begin{center}
\includegraphics[angle=0, viewport= 10 80 670 955, clip, width=\hsize]{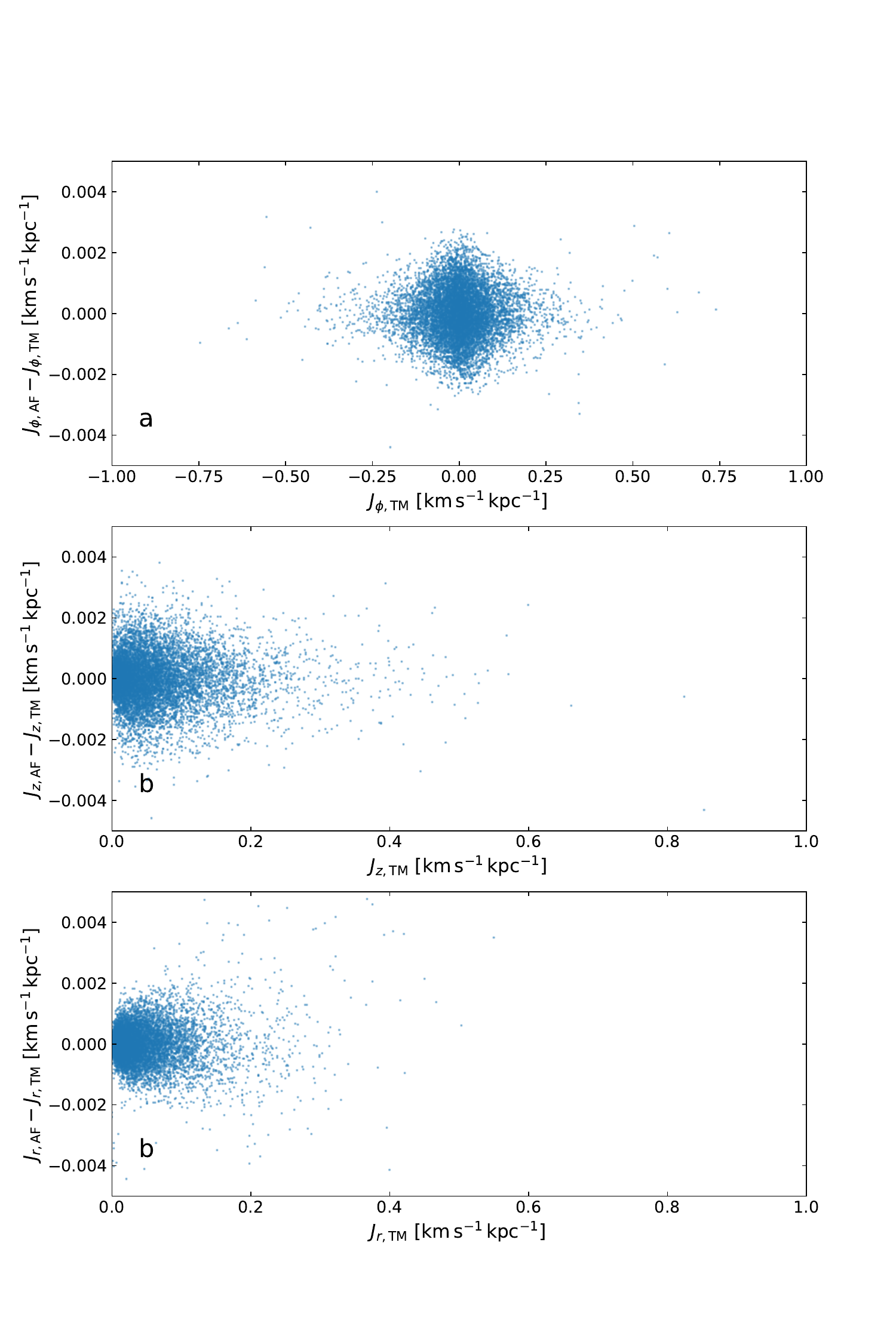}
\end{center}
\caption{Comparison between the actions calculated for a static Plummer model with the Torus Machinery (TM) in {\tt AGAMA}, and the actions derived from applying the \texttt{ACTIONFINDER} (AF) algorithm to 8 snapshots extracted from an N-body simulation of the same system.}
\label{fig:Nbody}
\end{figure}

For this test we generated a $10^6$ particle model taking a Plummer scale radius of $0.02\kpc$, and a total mass of $5\times10^5\msun$. The actions were calculated from the $(\pmb{x},\pmb{v})$ values using the spherical version of the action finder from {\tt AGAMA} (much more accurate than the Stäckel Fudge used for axisymmetric systems), in a potential corresponding to the Plummer sphere used to generate the initial conditions for the particles. The reader will have noticed that the parameters of our Plummer model make it resemble a relatively massive isolated globular cluster. We deliberately chose this example to emphasize the class of object that our algorithm is \emph{not} appropriate for. Real globular clusters are collisional systems, and undergo strong low-N interactions, invalidating the assumption of invariant actions. However, for the purpose of the test we avoid such strong interactions by applying a (Plummer kernel) force-softening to the particles during the simulation with a softening length of $0.001\kpc$. The particle system was integrated with the gyrfalcon N-body integrator \citep{2000ApJ...536L..39D} in the NEMO dynamics toolbox \citep{1995ASPC...77..398T}. We chose a total integration time of $0.07\Gyr$, setting the minimum timestep at $2^{-18}\Gyr$ (i.e. $\approx 3\,800\yr$), with 8 output snapshots at $0.01\Gyr$ intervals. The total integration time corresponds to approximately 3 dynamical times of this Plummer model at the half-mass radius.

The resulting difference between actions derived from the Torus Machinery and the present algorithm are displayed in Figure~\ref{fig:Nbody}. The mean absolute deviation of the differences between the two methods is $0.00055 \kms \kpc$ in $J_\phi$, $0.00048 \kms \kpc$ in $J_z$ and $0.00032 \kms \kpc$ in $J_r$. It is therefore probable that the \texttt{ACTIONFINDER} will be able to give useful approximations to the actions for certain dynamical N-body systems, but a more detailed exploration of its capabilities in this context is left to further dedicated works within our research program.

\section{Discussion and Conclusions}
\label{sec:Conclusions}

The present paper represents the first step of a research program aimed at using modern deep learning tools to perform the canonical transformation of positions and velocities to (and from) actions and angles variables from a sample of orbit-segments, and to find the underlying gravitational potential of the (simulated or real) system in an unsupervised way. Here, we do not deal with real data or complex simulated systems yet, but have constructed, and tested for simple configurations, a method to transform observable astrometric coordinates into action and angle coordinates with a scheme that is effectively a back-to-front version of that developed by \citet{McGill1990}, together with the canonical point-transformation improvement by \citet{1994MNRAS.268.1033K}. 

The deep neural nets used here are able to generate non-linear functions that are more flexible than a linear decomposition into Fourier coefficients, and this flexibility may make it easier to fit more general dynamical systems. A nice feature of our method is that it simultaneously uses {\it all} the data at its disposal to fit the generating functions of the canonical transformations. This contrasts to earlier methods where each orbit is fitted independently to derive the best set of Fourier coefficients for that orbit, bearing in mind that each such orbit requires many constraining data points due to the large number of Fourier coefficients that have to be fit. Moreover, once the transformation from positions and velocities is found, it is easy to inverse it in a supervised way. The main advantage of the method presented here, however, is that we do not need to know in advance the Hamiltonian or potential of the system under investigation. The algorithm finds it for us.

The data used in the tests presented here are noise free (apart from shot noise), and as such they do not portray a realistic picture of the limitations of the method as applied to real data. However, it is beyond the scope of the present contribution to attempt an exploration of the consequences of the observational limitations. With this caveat in mind, we are able to train the network to predict the actions from positions and velocities of halo-like orbits in a realistic axisymmetric Milky Way model, with an uncertainty of $\sim 0.6$\% of the action value of the Sun (compared to the input actions transformed into positions and velocities with the Torus Mapping), by training on only $N=8$ phase-space points along $S=128$ orbit segments, i.e. with only 1024 phase-space points. With $S=1024$ such orbit segments, an action uncertainty of $\sim 0.4$\% of the action value of the Sun can be attained. This is 5 times more accurate than the estimate of actions from positions and velocities with the St\"ackel fudge in {\tt AGAMA}. 

We also investigated the behaviour of the algorithm on a sample of disk-like orbits. Increasing the training sample to $S=8192$, we achieved an action accuracy of $0.67\kms \kpc$ (i.e. 0.03\% of the Sun's azimuthal action), again assuming that the Torus Machinery provides the ground-truth values. If we relax the requirement in the loss function for the returned angle variables to have the correct zero-points, we found that we could improve significantly the accuracy of the recovered actions to $0.088\kms \kpc$ using the same data set. It is possible that our algorithm could attain yet higher accuracy if it were provided a larger dataset (the depth of the neural net might also need to be increased).

We also directly compared the results of the Torus Machinery as implemented in \texttt{AGAMA} and the \texttt{ACTIONFINDER} on four test orbits defined by \citet{Sanders2016} (representing examples of a thin disk, a thick disk, a halo and a stream orbit). While it is difficult to gauge whether the comparison is fair (as the \texttt{ACTIONFINDER} trains its network based only on the data it was given along each orbit, yet at the same time it has to discover the geometry of the problem), the spread in the recovered actions (Figure~\ref{fig:test_orbits}) is very low, even lower than with the Torus Machinery as implemented in \texttt{AGAMA} in some cases.

For applications of the method to real data, we stress here that the algorithm requires the system under study to be sufficiently well sampled spatially over the galaxy so as to infer the global properties of the structure. For the Milky Way, the brightest nearby stars in future state-of-the-art Galactic datasets such as Gaia DR3 will have radial velocity accuracy of at best $0.1\kms$, while the 1\% distance uncertainty horizon lies at only $\approx 1\kpc$ from the Sun. Therefore, for our planned large-scale analysis of the Milky Way, there is no foreseeable need for the algorithm to return higher accuracy actions than we have demonstrated it is capable of.

For our method to be applicable to N-body simulations, the systems under study have to be in (or close to) equilibrium and the actions of the constituent particles need to exist and to be conserved. Exploring the limitations and pit-falls of this application of the algorithm would be a vast endeavor, far beyond the scope of the present contribution. Here we simply showed that if we use the particle positions and velocities at different timesteps from an N-body simulation of a Plummer model as inputs for the algorithm, the resulting actions are in good agreement with the values inferred for the N-body initial conditions, as calculated with the Torus Machinery.

Some studies might envisage analysing the dynamics of local stars through their differential motions, and the transformation of such relative motions into actions could require much higher accuracy than our method currently delivers. For such applications previous methods for calculating actions in exact potentials should be used. The reliability of the resultant findings would then depend on the applicability of the chosen potential model to the system under study, and the mismatch between that model and the actual system may end up dominating the error budget.

Our method may not be the ideal choice in situations where the exact potential is known in advance, such as for studies of orbits in analytic potential or density models. The algorithm was deliberately built to employ the flexibility of neural nets to adapt to the system the data are drawn from. When that flexibility is not needed, the method will incur a cost in the form of some coordinate transformation noise. That being said, the $G$ and $P$ generating functions can be forced to produce solutions consistent with a fixed potential model $\Phi_{\rm fixed}$: the only modification required to the algorithm is to add a loss term to Eqn.~\ref{eqn:loss} that compels Eqn.~\ref{eqn:J_dot} to hold. This loss term should be the mean absolute value of $dJ''_i/dt$, calculated from Eqn.~\ref{eqn:J_dot} (exactly as we implemented for the loss function of Eqn.~\ref{eqn:loss_acc}, but with $\ddot{x}$ given by $-\nabla \Phi_{\rm fixed}$ rather than by the acceleration network).

In principle, the present algorithm is not limited to axisymmetric systems, and given that the canonical coordinate transformation encoded in the generating function networks can be made arbitrarily complex by increasing the depth of the networks, one may be able to learn and model complex mass distributions given sufficient data and computational resources. However, the presence of chaotic dynamics in complex potentials gives cause for concern, since our method relies on the actions being conserved, which is not the case for chaotic orbits. It is thus likely that in such systems the method will only work for a subset of the orbits, while returning nonsensical results for the rest. We suspect that this could be turned from a bug into a feature of the method, as it may allow the present algorithm to be adapted into an automatic orbit classifier. For resonant orbits, it would be interesting to check whether the algorithm can find new orbital tori in each resonant trapping region, with their own system of angle-action variables \citep[e.g.,][]{Monari17, Binney18, Binney20a, Binney20b}.

We have attempted some very preliminary explorations of the method on a non-axisymmetric system by applying the algorithm to orbits integrated in a triaxial logarithmic halo model. We found that the present algorithm is not able to adapt automatically to fit all orbital families in a triaxial model simultaneously. The reason for this may be both because of the presence of chaotic orbits, and because the isochrone model is not a good starting point for the canonical transformation in a triaxial potential. Nevertheless, we were encouraged to find that we were able to fit a generating function that gave small scatter in $\pmb{J''}$ between points on the same orbit, by selecting the input sample of ``streams'' from high angular momentum tube orbits. In future work we will attempt to replace the isochrone model with a triaxial St\"ackel model \citep{deZeeuw85} as the ``toy'' starting point; it is plausible that this may provide the key to unlock unsupervised fitting in triaxial and more general systems. But we will also need to be able to identify chaotic orbits in the system.

The method currently relies on the fact that the potential is static. However, with the same caveats as above regarding chaos, it may be possible to generalize the algorithm so that motion is analysed in a rotating frame in which, for instance, a barred structure would appear static. This might allow the method to be used for analysing non-axisymmetric rotating systems as well. Of course, the application to real systems would have to consider carefully their actual complexities: in the case of the Milky Way evidence for a secularly slowing down bar has been provided by  \citet{2021MNRAS.500.4710C}, while it has also been recently shown that it may oscillate over time \citep{2020MNRAS.497..933H}. Whether such complications can be -- at least partially -- handled within the action-angle formalism, is of course an interesting question in itself largely independent of the method proposed in the present paper.

Enhancing the present algorithm to accept additional dynamical constraints will be relatively straightforward. For instance, rotation curve information can be trivially added without changing the code, simply by complementing the input sample with a number of kinematic points along circular in-plane orbits. Other dynamical information or priors can be provided by adding appropriate log-likelihood terms to the neural network's loss function. Furthermore, the uncertainties on the fitted model can be estimated by using the ``Dropout Layers'' to make alternative predictions \citep{2015arXiv150602142G}, providing similar information to what is traditionally derived from a (computationally much more expensive) Markov Chain Monte Carlo exploration.

The algorithm was deliberately built to accept astrometric data as inputs. Because {\it pytorch} processes data in parallel, this architecture makes it very simple to supply the network with multiple inputs for the same star, where the different instances could, for example, sample over the uncertainties in the astrometry, or scan over missing information in some input dimensions. Thus, if a star has a missing radial velocity measurement, one may attempt to find the radial velocity value that makes the derived actions for the star agree with those of the group. It is thus plausible that the present software can be adapted into a new method for detecting stellar streams, especially structures that do not possess any obvious spatial correlation.

The unsupervised learning technique developed here of building a network based on potentially complex corrections to simple analytic models and applying the physical constraints as loss function terms may have substantially wider applications. The {\it pytorch} tensor structure is particularly powerful for this purpose, with its ability to implement automatically the derivatives of the analytic model and the correction function. As we have seen, this provides an easy means to incorporate differential equations into a network.

A simplified demonstration version of the code, along with a sample of test data from the \citet{1998MNRAS.294..429D} potential model `1' is available on Zenodo\footnote{\texttt{https://doi.org/10.5281/zenodo.4664482}} (DOI: 10.5281/zenodo.4664482) and GitHub\footnote{\texttt{https://github.com/RodrigoIbata/ActionFinder}}.

\acknowledgments

RI, BF, and GM acknowledge funding from the Agence Nationale de la Recherche (ANR project ANR-18-CE31-0006, ANR-18-CE31-0017 and ANR-19-CE31-0017), from CNRS/INSU through the Programme National Galaxies et Cosmologie, and from the European Research Council (ERC) under the European Unions Horizon 2020 research and innovation programme (grant agreement No. 834148).

\bibliography{ACTIONFINDER}
\bibliographystyle{aasjournal}

\section*{Appendix}
\label{sec:Appendix}

\begin{python}[emphstyle=\textcolor{magenta}, caption={Pytorch pseudocode for the ActionFinder module.}, emph={forward, ActionFinder},label={list:DECODE}]

class ActionFinder(torch.nn.Module):
    """
    Inputs: 
           d,l,b,vh,mu_l,mu_b astrometric phase space coordinates
    Output: 
           Jdd_vals - the J'' values
           Tdd_vals - the theta'' values
           loss_Jdd_spread - loss due to the spread of the J'' values
    """
    def __init__(self,verbose=False):
        super().__init__()

        self.conv_input2xv = conv_Input2xv(verbose=verbose)
        self.isochrone_analytic = Isochrone_Analytic(verbose=verbose)
        self.encoder_GF_G = Encoder_GF_G(verbose=verbose)
        self.encoder_GF_P = Encoder_GF_P(verbose=verbose)

    def forward(self,inputs): #               
    
        inputs_xv = self.conv_input2xv(inputs)        # convert d,l,b,vh,mu_l,mu_b to xv
        
        # (J,Theta) in isochrone toy model 
        J_iso,  T_iso = self.isochrone_analytic(inputs_xv, M, b)

        # We now iterate to find J'. Start with J' = mean(J_iso for each stream)
        Jd_mean = torch.mean( J_iso, dim=1, keepdim=True)

        for iter in range(iter_max):
            Jd_trial = Jd_mean.clone()                # trial J'
            Jd_mean_fill = torch.cat([Jd_mean]*NStars,dim=1) # expand mean value to fill NStars dimension

            TJd  = torch.cat( [T_iso,Jd_mean_fill],dim=-1 ) # (T,J')
            GF_G = self.encoder_GF_G(TJd)             # generating function G
            # J-J' = d(G)/d(T):
            JmJd = jacobian(T_iso,GF_G)[:,:,0,:].data # remove from computational graph
            GF_G.detach()                             # remove from computational graph

            # update 
            Jd_mean = torch.mean( J_iso - JmJd, dim=1, keepdim=True)

            if ( torch.max( torch.abs(Jd_trial - Jd_mean)) < 5.e-5 and # test for acceptable convergence
                 torch.mean(torch.abs(Jd_trial - Jd_mean)) < 1.e-6):
                break
                
                
        # Find Jd_mean once more, but now retain the computational graph 
        Jd_mean_fill = torch.cat([Jd_mean]*NStars,dim=1)
        TJd  = torch.cat( [T_iso,Jd_mean_fill],dim=-1 )
        GF_G = self.encoder_GF_G(TJd)
        JmJd = jacobian(T_iso,GF_G)[:,:,0,:]
        Jd_vals = J_iso - JmJd                       # J' given by generating function G
        Jd_mean = torch.mean( Jd_vals, dim=1, keepdim=True)

        # Calculate T' from generating function derivative
        Jd_mean_fill = torch.cat([Jd_mean]*NStars,dim=1)
        TJd  = torch.cat( [T_iso,Jd_mean_fill],dim=-1 )
        GF_G = self.encoder_GF_G(TJd)

        # T' = T + d(G)/d(J'):
        Td_vals = (T_iso + jacobian(Jd_mean_fill,GF_G)[:,:,0,:]) % (2*torch.pi)

        
        # Add in the extra freedom of a point-transformation
        Tdd_vals = self.encoder_GF_P( Td_vals )      # final T'' from generating function P
        dTdd_dTd = jacobian(Td_vals,Tdd_vals)        # d(T'')/d(T')
        dTd_dTdd = torch.inverse( dTdd_dTd )         # d(T')/d(T'')
        # J_i'' = dT'/dT_i'' . J'
        Jdd_vals = torch.einsum('bsji,bsj->bsi',dTd_dTdd,Jd_vals) # final J'' values

        Jdd_mean = torch.mean( Jdd_vals, dim=1, keepdim=True)
        Jdd_spread = Jdd_mean - Jdd_vals
        
        loss_Jdd_spread = torch.mean( torch.abs( Jdd_spread ) )
        
        return Jdd_vals, Tdd_vals, loss_Jdd_spread
\end{python}

\begin{python}[emphstyle=\textcolor{magenta}, caption={Pytorch pseudocode for the unit ResDense block.}, emph={forward, ResDense_block},label={list:UNIT_BLOCK}]

class ResDense_block(torch.nn.Module):
    """
    The ResDense unit block.
    This layer does not change the number of features. 
    """
    
    def __init__(self, nunits):
        super().__init__()
        """
        Declaration of the layers that will be used:
        nunits is the number of features.
        """
        
        self.nunits = nunits
            
        self.dense1 = torch.nn.Linear(in_features=self.nunits, out_features= self.nunits, bias=False)
        self.WN1    = torch.nn.utils.weight_norm(self.dense1)
        self.dense2 = torch.nn.Linear(in_features=self.nunits, out_features= self.nunits)
        self.WN2    = torch.nn.utils.weight_norm(self.dense2)
    
    def forward(self, input):
        """
        The computational graph (how the above layers are used):
        """
        
        xx = torch.relu(input)
        xx = self.dense1(xx)
        xx = self.WN1(xx)

        xx = torch.relu(xx)
        xx = self.dense2(xx)
        xx = self.WN2(xx)

        output = xx + input
            
        return output
\end{python}

\end{document}